# Anisotropy of acoustooptic figure of merit in optically isotropic media

Oksana Mys, Myroslav Kostyrko, Mykola Smyk, Oleh Krupych and Rostyslav Vlokh *

*Institute of Physical Optics, 23 Dragomanov Street, 79005 Lviv, Ukraine*
*Corresponding author: vlokh@ifo.lviv.ua*

We suggest a new approach for analyzing spatial anisotropy of acoustooptic figure of merit (AOFM). The relations for the effective elastooptic coefficients and the AOFM are derived for all possible types of acoustooptic (AO) interactions in optically isotropic media, including non-solid-state and solid-state amorphous media, and crystals belonging to the cubic system. Our approach allows for finding the optimal geometries of AO interactions characterized by the highest AOFM for a given material. The analysis is carried out on the examples of cubic KBr and $KAl(SO_4)_2 \times 12H_2O$ crystals which represent different subgroups of the cubic symmetry class.

http://dx.doi.org/10.1364/AO.99.099999

## 1. Introduction

Among the parametric optical effects induced by external fields, such as electro-, magneto-, piezo- and acoustooptic (AO) ones [1, 2], AO diffraction represents maybe the most frequently used effect. This phenomenon is utilized for deflecting and modulating of light, scanning of the optical beams, RF spectrum analyzing, Q-switching, etc. [3-6]. It is well known that efficiency of the AO diffraction is defined by the AO figure of merit (AOFM):

$$M_2 = \frac{n^6 p_{ef}^2}{\rho v^3}, \qquad (1)$$

where $n$, $p_{ef}$ and $\rho$ denote respectively the optical refractive index, the effective elastooptic coefficient (EEC) and the density of an optical material, and $v$ the velocity of the acoustic wave (AW). Although the $M_2$ parameter is scalar, it can reveal notable spatial anisotropy. This anisotropy is mainly influenced by the elastooptic tensor and the tensor of elastic stiffness coefficients, on the basis of which the AW velocities are determined. Anisotropy of the refractive indices contributes to the total spatial dependence of $M_2$, too. Thus, the AO efficiency simultaneously depends on a number of constitutive coefficients of a given optical material. Searching for efficient materials for the AO diffraction is of a primary importance. Moreover, development of the methods for analyzing the anisotropy of AO efficiency for different geometries of AO interactions can solve a lot of materials science problems, since the anisotropy of the $M_2$ coefficient can yield in increasing AO efficiency for both well-known and new materials. As shown in our recent works [7, 8], the anisotropy of AW velocities often plays a major role in the anisotropy of AOFM. In fact, it can result in extremely high $M_2$ coefficient, e.g., for one of the best AO materials, $TeO_2$ crystal [9, 10].

It is obvious that the EEC and its anisotropy also contribute significantly to the AO efficiency. In particular, this contribution can become notable for the media where the AW velocities are almost isotropic (see, e.g., [11]). Unfortunately, no general method for analyzing the anisotropy of AO efficiency has been developed up to now. To our best knowledge, there have been a few relevant attempts, which are based on the anisotropies of the EEC [12] and the AW velocities [2]. A combined method has been suggested in Ref. [13, 14]. However, it is mainly declarative and represents in fact an intention for developing a method for the analysis of AO efficiency anisotropy. In addition, a procedure of transformation of the elastooptic tensor in the coordinate systems coupled with polarization direction of the incident optical wave, which has been used in Ref. [13, 14], does not solve the problem. Such a procedure can be used, for example, when analyzing the longitudinal piezoelectric effect (see, e.g., [15]) but not in the case of AO interactions which are more complicated and include a lot of parameters. Thus, the aim of the present work is to start development of a general method for analyzing the anisotropy of the $M_2$ coefficient. It is concerned with optically isotropic media, including isotropic gaseous and liquid materials, isotropic solid-state media and crystals of the cubic system.

## 2. Results of analysis

We begin our analysis from the simplest case of isotropic gaseous or liquid media which are optically and acoustically isotropic. Only volume AWs (or density waves) can propagate in liquids and gases. Propagation of transverse shear waves is not supported in such media. We will consider optically isotropic solid-state media such as glasses, etc. Here the eigenwaves can be longitudinal or transverse AWs. The velocity surfaces for these waves are represented by spheres and, moreover, the velocities of the transverse waves with mutually orthogonal polarizations are the same. Our next step will be the analysis of AO interactions in the cubic crystals, which are isotropic optically but anisotropic acoustically. The analysis for optically anisotropic crystals belonging to the middle and lower crystallographic systems (i.e., for optically uniaxial and biaxial crystals) will be presented in our forthcoming works. Such a sequence of subjects under analysis seems to be logical since each next step makes the consideration more complicated.

### 2.1. Isotropic liquid and gaseous media

Gaseous media are characterized by relatively high AOFMs due to their low AW velocities. For instance, the figures of merit for such gases as nitrogen and oxygen are equal respectively to $680\times10^{-15} s^3/kg$ and $589\times10^{-15} s^3/kg$ at the pressure 100 bar and the room temperature [16]. Xenon is characterized by the $M_2$ coefficient equal to $\sim 5130\times10^{-15} s^3/kg$ at the pressure of 50 bar and the room temperature, while SF$_6$ reveals an extremely high AOFM ($15400\times10^{-15} s^3/kg$ at the same temperature and the pressure of 20 bar [16]). Notice also that the AOFM for the liquid xenon (at 258 K and the hydrostatic pressure 29 bar) increases approximately two times when compared with the corresponding gas, and is equal to $11630\times10^{-15} s^3/kg$ [16].

The velocity of propagation of the volume waves in the isotropic liquid and gaseous media is given by

$$v = \sqrt{\frac{b}{r}} = \sqrt{\frac{C_{11}+2C_{12}}{3r}} \ , \qquad (2)$$

where $b = (C_{11}+2C_{12})/3$ is the bulk elastic module [2], $r$ denotes the density, $C_{ij}$ are elastic stiffness coefficients. Let the polarization of the incident optical wave be parallel to $Z$ axis and the AW propagate along $Y$ axis (see Fig. 1).

Since the optical impermeability tensor $B_k$ is directly coupled to the strain tensor $e_j$ via elastooptic effect further we will consider the increment $\Delta B_k = p_{kj} e_j$, which can be included into the relation for dielectric permittivity as $e_k = e_{(0)k} - e_{(0)k}^2 \Delta B_k$ ($p_{kj}$ are elastooptic tensor coefficients, $e_{(0)k}$ is the initial value of dielectric permitivity). The electric field of the diffracted wave can be represented as

$$E_3 = \Delta B_3 D_3 = (p_{12}e_1 + p_{12}e_2 + p_{11}e_3)D_3 = (2p_{12}+p_{11})eD_3 , \qquad (3)$$

where $D_3$ is the electric induction component of the incident wave, $\Delta B_3$ the increment of the optical impermeability tensor component, and $e = e_1 = e_2 = e_3$ are strain tensor components which contribute at the propagation of the bulk wave. Notice, here we consider the Bragg diffraction which can be reached by the appropriate choice of the interaction geometry and length. Then the refractive index change induced by the AW in the direction which is parallel to the polarization of the incident wave is as follows:

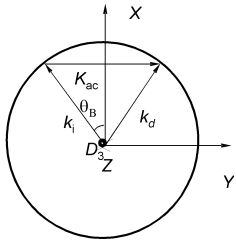

Fig. 1. Scheme of AO interaction in isotropic liquid or gaseous media: $X$, $Y$, $Z$ are axes of laboratory coordinate system, $K_{ac}$ the AW vector, $k_i$ and $k_d$ the wave vectors respectively of the incident and diffracted optical waves, and $q_B$ the Bragg angle.

$$\Delta n_3 = n^3(2p_{12}+p_{11})e/2 , \qquad (4)$$

and the EEC is given by

$$p_{ef} = (2p_{12}+p_{11}) . \qquad (5)$$

Finally, the AOFM reads as

$$M_2 = \frac{n^6(2p_{12}+p_{11})^2}{rv^3} = \frac{3n^6(2p_{12}+p_{11})^2\sqrt{3r}}{\left(\sqrt{(2C_{12}+C_{11})}\right)^3} . \qquad (6)$$

As a result, the liquids and gases are characterized by a single value of AOFM, which has no dependence on the interaction geometry – neither the AW propagation direction nor the Bragg angle, because of rotational invariance of all the quantities included into Eq. (6).

### 2.2. Isotropic solid-state media

Amorphous isotropic glasses represent convenient and efficient materials for AO applications. For example, chalcogenide glasses have high values of the AOFM exceeding $10^{-12} s^3/kg$ and are transparent in the far IR range [17]. Although the $M_2$ coefficient for the fused silica is relatively small, this glass is transparent in the deep UV range [8]. The borate glasses are resistant to powerful laser radiation and transparent in the UV range, and manifest high enough figures of merit [18].

As mentioned above, the two types of AWs can propagate in the isotropic solid-state media, the longitudinal and transverse ones. The velocities of these waves do not depend on the direction of propagation. In addition, the velocity of the transverse waves does not depend on their polarization. These conclusions follow from the Christoffel equation,

$$C_{mnop}K_n K_o p_p = rv^2 p_m , \qquad (7)$$

Where $C_{mnop} = C_{ij}$ at $m, n, o, p = 1,…3$ and $i, j = 1, …, 6$, $K_n$, $K_o$ are the components of the AW vector and $p_n$, $p_m$ the components of the AW polarization [2, 19, 20]. There are only two velocity solutions for the case of isotropic solid-state media:

$$v^2 = (C_{11}-C_{12})/(2r) \qquad (8)$$

for the transverse wave and

$$v^2 = C_{11}/r \qquad (9)$$

for the longitudinal one. The existence of the two types of acoustic eigenwaves leads to increasing number of possible types of the AO interactions and so of the corresponding parameters of the AOFM. Here one has to distinguish among the three types of interaction: (I) AO interaction of the longitudinal AW (e.g., $v_{11}$) propagating along the $X$ axis with the incident optical wave, of which electric induction vector is parallel to the $Y$ axis (see Fig. 2,a); (II) AO interaction of the longitudinal AW $v_{11}$ propagating, e.g., along the $X$ axis with the incident optical wave, of which electric induction lies in the $XZ$ plane at the angle $q_B$ with respect to the $X$ axis ($D_3 = D\sin q_B$ and $D_1 = D\cos q_B$ – see Fig. 2,b); and (III) AO interaction of the transverse AW $v_{13}$ propagating along the $X$ axis and polarized along the $Z$ axis with the incident optical wave, of which polarization vector lies in the $XZ$ plane at the angle $q_B$ with respect to the $X$ axis ($D_3 = D\sin q_B$ and $D_1 = D\cos q_B$ – see Fig. 2,c). Notice that a fourth hypothetically possible type of interaction of the transverse AW $v_{13}$ with the incident optical wave polarized parallel to the $Y$ axis cannot be implemented since the relevant elastooptic coefficient $p_{25}$ is equal to zero in all isotropic solid-state media. Here we adopt the notation for AW velocities as $v_{ij}$, where index $i$ indicates the direction of propagation and index $j$ – the direction of polarization (see e.g. [21]).

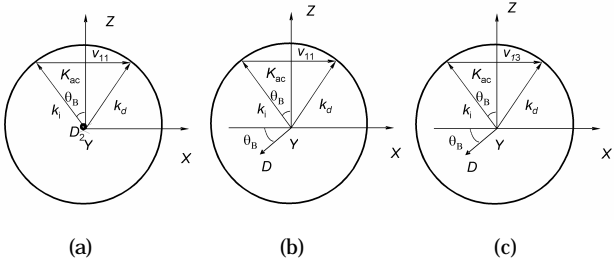

Fig. 2. Three types of AO interactions possible in the isotropic amorphous solids (see explanations in the text).

In the case (I) the relations for the electric field of the diffracted optical wave and the acoustically induced increment of the refractive index are as follows:

$$E_2 = \Delta B_2 D_2 = p_{12} e_1 D_2, \qquad (10)$$

$$\Delta n_2 = n^3 p_{12} e_1 / 2, \qquad (11)$$

where $p_{ef} = p_{12}$. Then the figure of merit does not depend on the propagation directions of the acoustic and optical waves:

$$M_2^{(I)} = n^6 p_{12}^2 r^{-1} v_{11}^{-3}. \qquad (12)$$

In the case (II) we have $E_1 = \Delta B_1 D_1 = p_1 e_1 D_1$ and $E_3 = \Delta B_3 D_3 = p_{12} e_1 D_3$. The changed refractive index for the direction of incident light wave polarization is

$$n_1'^* = n - n^3 (p_{12} \sin^2 q_B + p_{11} \cos^2 q_B) e_1 / 2. \qquad (13)$$

Then the EEC reads as

$$p_{ef} = (p_{12} \sin^2 q_B + p_{11} \cos^2 q_B), \qquad (14)$$

while the relation for the AOFM becomes as follows:

$$M_2^{(II)} = n^6 \left( p_{12} \sin^2 q_B + p_{11} \cos^2 q_B \right)^2 r^{-1} v_{11}^{-3}. \qquad (15)$$

It is seen that the AOFM depends on the Bragg angle, i.e. on the AW frequency. In the high-frequency range, the influence of coefficient $p_{12}$ on $M_2$ can become significant, though only $p_{11}$ will play a major part for lower frequencies.

In the case (III) the transverse wave can induce the component $e_5$ of the strain tensor. The electric field components of the diffracted optical wave are defined by the relations $E_1 = \Delta B_5 D_3 = p_{44} e_5 D_3$ and $E_3 = \Delta B_5 D_1 = p_{44} e_5 D_1$ (for isotropic media $p_{44} = p_{55} = (p_{11} - p_{12})/2$), whereas the refractive index change for the direction of incident light wave polarization is given by $\Delta n = n^3 p_{44} e_5 \sin 2q_B / 2$, where $p_{ef} = p_{44} \sin 2q_B$. Therefore the AOFM is determined by the relation

$$M_2^{(III)} = n^6 \left( p_{44} \sin 2q_B \right)^2 r^{-1} v_{13}^{-3}. \qquad (16)$$

In other words, the figure of merit depends strongly on the Bragg angle. Since the Bragg angle is very small, the AOFM for this type of interaction also remains small. In fact, this type of interaction corresponds to the condition under which the AW polarization is almost parallel to the incident light propagation direction.

Let us summarize the results of our analysis for the AOFM of the isotropic media. In all the cases considered, the figure of merit manifests no anisotropy, being dependent on the Bragg angle. The liquid and gaseous media are characterized by a single value of the figure of merit, while the isotropic solid-state media can be described by the three different AOFMs, which correspond to the interaction between different acoustic and optical waves. Let us remind the AO interaction in borate glasses as example [18]. It should be noted that AO interaction with the transverse AW which in [18] is characterized by high AOFM due to results of our present consideration can not be realized or will lead to quite low value of AOFM. While interaction with longitudinal AW correspond to (I) and (II) types of interaction.

### 2.3. Cubic crystals

Among the cubic crystals, one can find a lot of efficient AO materials. For instance, the AOFM is equal to $840 \times 10^{-15}\, s^3/kg$ for Ge crystals, $44.6 \times 10^{-15}\, s^3/kg$ for GaP, and $104 \times 10^{-15}\, s^3/kg$ for GaAs [10]. In addition, some of the cubic crystals (e.g., KBr) reveal a notable anisotropy of AW velocities which can be efficiently used while optimizing the geometry of AO interactions.

The forms of the elastic stiffness tensors for all the crystals belonging to the cubic system are the same, while the tensor of elastooptic coefficients for the crystals of symmetry groups m3m, 432 and $\bar{4}$3m differs from that for the groups 23 and m3. This tensor includes three independent coefficients $p_{11}$, $p_{12}$ and $p_{44} \neq (p_{11} - p_{12})/2$ for the first subgroup and four coefficients for the second subgroup, for which the relationship $p_{21} \neq p_{12}$ holds true. The elastic stiffness tensor for the crystals belonging to the cubic system consists of three independent coefficients $C_{11}$, $C_{12}$ and $C_{44} \neq (C_{11} - C_{12})/2$, thus removing degeneracy for the transverse waves with mutually orthogonal polarizations, in contrast to amorphous media. These velocities become different for the so-called quasi-transverse waves QT$_1$ and QT$_2$:

$$v_{QT_1}^2 = \frac{1}{2r} \left[ C_{11} + C_{44} - \sqrt{\begin{array}{l}(C_{11} - C_{44})^2 \cos^2 2\Theta + \\ + \sin^2 2\Theta (C_{12} + C_{44})^2\end{array}} \right], \qquad (17)$$

$$v_{QT_2}^2 = C_{44}/r. \qquad (18)$$

The relation for the quasi-longitudinal wave velocity looks as follows:

$$v_{QL}^2 = \frac{1}{2r} \left[ C_{11} + C_{44} + \sqrt{\begin{array}{l}(C_{11} - C_{44})^2 \cos^2 2\Theta + \\ + \sin^2 2\Theta (C_{12} + C_{44})^2\end{array}} \right], \qquad (19)$$

where $\Theta$ is the angle between the AW vector and the $X$ axis in the $XZ$ plane. This is the angle of rotation of the AW vector around the $Y$ axis and also the angle of rotation of vector diagram of AO interaction around the $Y$ axis (see Fig. 3). Notice that here the coordinate system $XYZ$ coincides with the crystallographic system $abc$. As seen from Eqs. (17)–(19), the velocities of the three acoustic eigenwaves depend on the AW vector direction. In fact, Eqs. (17)–(19) describe the dependences of velocities of the acoustic eigenwaves propagating in the cubic crystals on the wave vector direction in the $XZ$ plane.

Our next step is elucidating the dependence of the EEC on the direction of AW vector in the $XZ$ plane. Notice that the EECs differ for different types of AO interactions and different elastooptic subgroups of the cubic system. The types of AO interactions for the cubic crystals are similar to those listed above for the isotropic amorphous solid-state media (see Fig. 2). We will start our analysis with the subgroup that embraces the symmetry groups m3m, $\bar{4}3m$ and 432.

The three types of interactions mentioned above include: (I) AO interaction of the longitudinal AW $v_{11} = v_{QL}$ propagating along the $X$ axis with the incident optical wave, of which electric induction vector is parallel to the $Y$ axis; (II) AO interaction of the longitudinal AW $v_{11} = v_{QL}$ propagating, e.g., along the $X$ axis with the incident optical wave, of which electric induction lies in the $XZ$ plane at the angle $q_B$ with respect to the $X$ axis ($D_3 = D\sin q_B$ and $D_1 = D\cos q_B$); and (III) AO interaction of transverse AW $v_{13} = v_{QT_1}$ propagating along the $X$ axis and polarized along the $Z$ axis with the incident optical wave, of which polarization vector lies in the $XZ$ plane at the angle $q_B$ with respect to the $X$ axis ($D_3 = D\sin q_B$ and $D_1 = D\cos q_B$ – see Fig. 2,c). Note that the additional type (IV) of interaction of the transverse AW $v_{13}$ with the incident optical wave polarized parallel to the $Y$ axis is possible in the cubic crystals. The same is true of the types (V) and (VI) of AO interactions, i.e. the interactions with the $QT_2$ wave are possible, too. It is truth that these types of interaction cannot be implemented in the $XZ$ plane since the appropriate elastooptic coefficients $p_{16}$ and $p_{26}$ are equal to zero. However, these coefficients become nonzero if one considers the AO interaction occurring in the $X'Z$ planes formed by rotating the interaction plane around the $Z$ axis by some angle $j$ (see Fig. 4).

For the type (I) of interaction, the EEC remains unchangeable after rotation of AW vector in the $XZ$ plane and is equal to $p_{12}$. Notice that rotation of the AW vector in the $XZ$ plane is equivalent to rewriting of the strain tensor that includes only one component $e_1$ in the coordinate system rotated around the $Y$ axis by the angle $\Theta$. Then the three tensor components appear:

$$\tilde{e}_1 = e_1 \cos^2 \Theta; \quad \tilde{e}_3 = e_1 \sin^2 \Theta; \quad \tilde{e}_5 = e_1 \sin\Theta \cos\Theta. \quad (20)$$

As a result, the EEC is given by

$$p_{ef} = p_{12} \cos^2 \Theta + p_{13} \sin^2 \Theta = p_{12}, \quad (21)$$

since we have $p_{13} = p_{12}$. The anisotropy of the AOFM is defined solely by the anisotropy of QL wave velocity in the $XZ$ plane, i.e. by the dependence $v_{QL}(\Theta)$. Under rotation of the interaction plane around the $Z$ axis by the angle $j$, the EEC will change to

$$p_{ef}^{(I)} = p'_{12} \cos^2 \Theta + p'_{13} \sin^2 \Theta \Big|_{p'_{13}=p_{12}} = $$
$$= p_{12} \sin^2 \Theta + \begin{bmatrix} p_{12}(\cos^4 j + \sin^4 j) + \\ + \sin^2 2j (p_{11} - 2p_{44})/2 \end{bmatrix} \cos^2 \Theta, \quad (22)$$

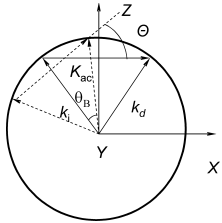

Fig. 3. Vector diagram of AO interaction and the same diagram rotated by the angle $\Theta$ around the $Y$ axis.

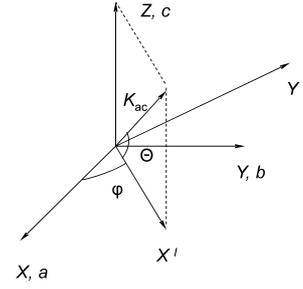

Fig. 4. Crystallographic coordinate system coupled with the Cartesian system $XYZ$, and a new coordinate system $X'Y'Z$ coupled with the plane of AO interaction $X'Z$.

where $p'_{kj}$ are the components of elastooptic tensor rewritten in the coordinate system $X'Y'Z$ rotated by the angle $j$ around the $Z$ axis with respect to the $XYZ$ system.

In the latter case one has to take into account the changes in the AW velocity and the cross section of the AW velocity surface by the $X'Z$ plane, i.e. the dependences $v_{QL}(\Theta,j)$. This can be made by solving the determinant based on the Christoffel tensor:

$$N_{mp} = C_{mnop} K_n K_o. \quad (23)$$

Here the components of the elastic stiffness tensor $C_{mnop}$ depend on the angle $j$ and the AW vector components $K_n$ and $K_o$ on the angle $\Theta$. As a result, a cubic equation has to be solved, of which solutions can be obtained using the standard numeric techniques. For the type (I) of AO interaction, the relation for the AOFM is as follows:

$$M_2^{(I)} = n^6 \left[ p_{ef}^{(I)} \right]^2 r^{-1} \left[ v_{QL}(\Theta,j) \right]^{-3}. \quad (24)$$

We have simulated the anisotropy of AOFM for this cubic subgroup on the example of KBr belonging to the symmetry group m3m [8]. All the necessary data are available in the literature: $r = 2750 \text{kg/m}^3$, $C_{11} = 34.6\text{GPa}$, $C_{12} = 5.6\text{GPa}$, $C_{44} = 5.15\text{GPa}$, $n = 1.559$, $p_{11} = 0.241$, $p_{12} = 0.191$, and $p_{44} = -0.023$ (for the light wavelength $l = 589\text{nm}$). The results of simulations are displayed in Fig. 5. The anisotropy of the AOFM is mainly governed by anisotropy of the longitudinal AW velocity (notice that Fig. 5,c shows the acoustic slowness $v_{11}^{-1}$ for more clearness, since the AW velocity enters into denominator of the AOFM). The parameter $M_2^{(I)}$ reaches its maximum $8.67 \times 10^{-15} \text{s}^3/\text{kg}$ at $\Theta = 42\deg$ and $j = 45\deg$. The anisotropy of $p_{ef}$ also changes $M_2^{(I)}$, in particular for the AW propagation along the $Z$ and $X'$ axes. However, the latter anisotropy factor is quite small for the crystals considered here.

Let us proceed to the type (II) of AO interactions, i.e. the interaction of the longitudinal AW $v_{11} = v_{QL}$ propagating along the $X$ axis with the optical wave, whose electric induction lies in the $XZ$ plane at the angle $q_B$ with respect to the $X$ axis ($D_3 = D\sin q_B$ and $D_1 = D\cos q_B$). The strain tensor components involved are defined by Eqs. (20). The EEC in the arbitrary $X'Z$ plane of interaction is as follows:

$$p_{ef}^{(II)} = \left\{ \begin{bmatrix} p_{11}(\cos^4 j + \sin^4 j) + \\ \sin^2 2j (p_{12} + 2p_{44})/2 \end{bmatrix} \cos^2 \Theta + p_{12} \sin^2 \Theta \right\} \times$$
$$\times \cos^2(q_B + \Theta) - \left[ p_{44} \sin 2(q_B + \Theta) \sin 2\Theta \right]/2 + \quad . \quad (25)$$
$$+ \sin^2(q_B + \Theta)\left( p_{12} \cos^2 \Theta + p_{11} \sin^2 \Theta \right)$$

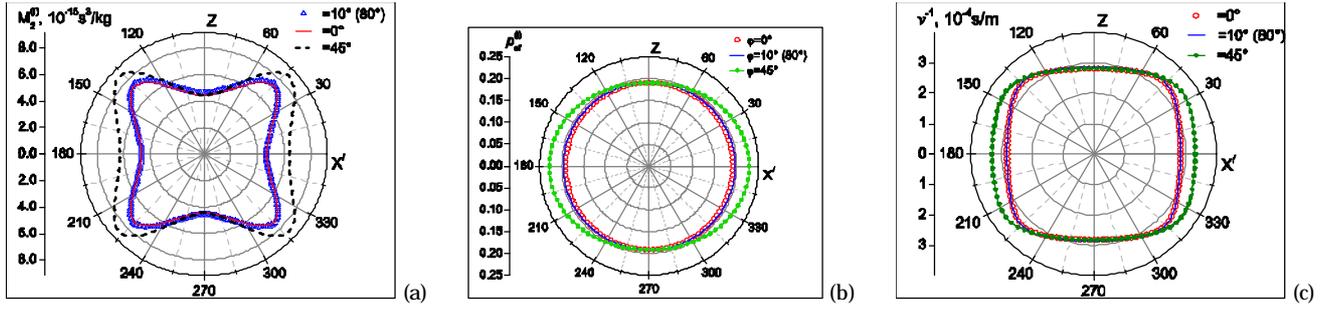

Fig. 5. Dependences of $M_2^{(I)}$ coefficient (a), EEC (b) and acoustic slowness (c) on the direction of AW vector (angle $\Theta$) at different orientations of the interaction plane (angle $\varphi$).

Then the AOFM reads as

$$M_2^{(II)} = n^6 \left[ p_{ef}^{(II)} \right]^2 \rho^{-1} \left[ v_{QL}(\Theta,\varphi) \right]^{-3}. \qquad (26)$$

It is clearly seen that the AOFM depends on the Bragg angle. This dependence is presented in Fig. 6,a for the case of KBr crystals. The AOFM remains in fact unchangeable for small Bragg angles (0.24–4 deg) though it changes at large angles (45 deg). The latter case corresponds to extremely high AW frequencies which are not used in practice for AO interactions. A decrease in the AOFM at larger Bragg angles is caused by decreasing EEC (see Fig. 6,b). The other peculiarity is that the anisotropy of AOFM results only from the anisotropy of AW slowness (see Fig. 5,c), because the EEC does not depend on the AW vector direction in the $XZ$ plane.

In our further analysis, we will take the value of the Bragg angle equal to 4 deg. It is seen from Fig. 7,a that the maximal AOFM is reached at $\varphi = 0 \deg$, i.e. in the $XZ$ plane and for the AW vector parallel to the bisector of the $Z$ and $X$ axes. Under these conditions we obtain $M_2^{(II)} = 12.8 \times 10^{-15} \text{s}^3/\text{kg}$. Similar to the case of AO interaction in the $XZ$ plane, the main contribution to the anisotropy of $M_2^{(II)}$ coefficient is caused by the anisotropy of AW slowness in the $X'Z$ plane, again because of a weak dependence of EEC on the direction of AW vector.

Now let us concentrate on the type (III) of AO interaction of the optical wave, whose polarization vector lies in the $XZ$ plane at the angle $\theta_B$ with respect to the $X$ axis, with the transverse AW $v_{13} = v_{QT_1}$ propagating along the $X$ axis and polarized along the $Z$ axis. Here we have $D_3 = D \sin\theta_B$ and $D_1 = D \cos\theta_B$, and the strain tensor caused by the AW contains a single component $e_5$. When the AW vector direction changes in the $XZ$ plane by the angle $\Theta$, the components of the strain tensor may be written as

$$\tilde{e}_1 = e_5 \sin 2\Theta; \quad \tilde{e}_3 = -e_5 \sin 2\Theta; \quad \tilde{e}_5 = e_5 \cos 2\Theta. \qquad (27)$$

The dependence of the EEC on the AW vector direction in the

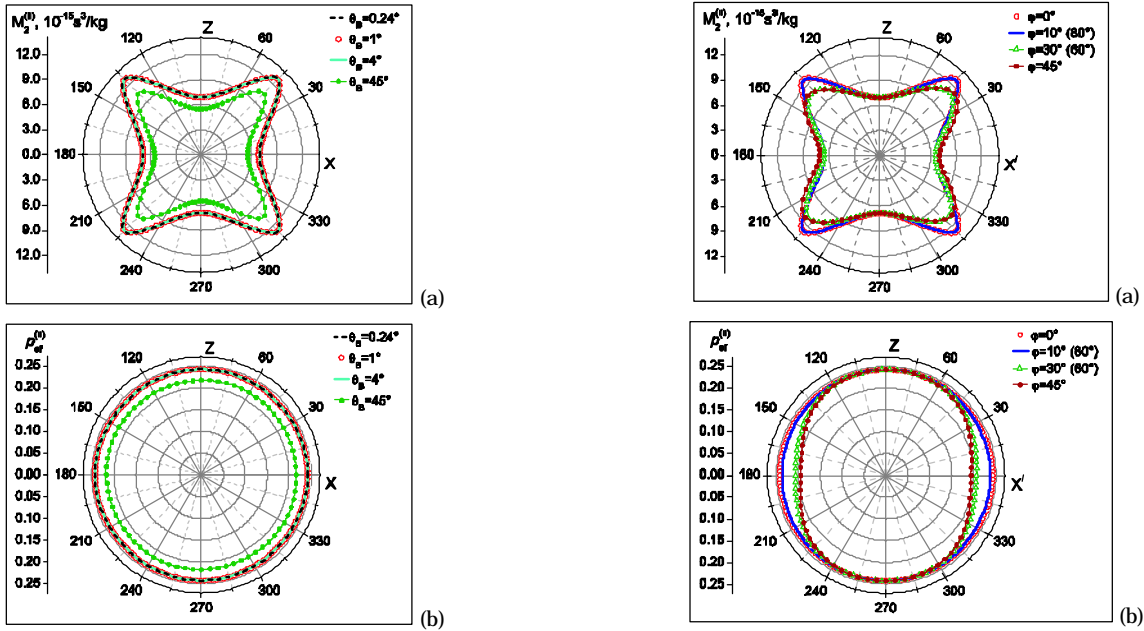

Fig. 6. Dependences of AOFM $M_2^{(II)}$ (a) and EEC (b) on the AW vector direction in the $XZ$ plane at different Bragg angles, as simulated for KBr crystals.

Fig. 7. Dependences of $M_2^{(II)}$ coefficient (a) and EEC (b) on the direction of AW vector (angle $\Theta$) at different orientations of the interaction plane (angle $\varphi$).

arbitrary $X'Z$ plane is given by

$$p_{ef}^{(III)} = \cos^2(\mathbf{q}_B + \Theta)\sin 2\Theta \times$$
$$\times \begin{bmatrix} \sin^2 2\mathbf{j}\,(p_{12}+2p_{44})/2 + \\ +p_{11}(\cos^4\mathbf{j}+\sin^4\mathbf{j}) - p_{12} \end{bmatrix} + \quad (28)$$
$$+\sin^2(\mathbf{q}_B+\Theta)\sin 2\Theta(p_{12}-p_{11}) +$$
$$+p_{44}\sin 2(\mathbf{q}_B+\Theta)\cos 2\Theta$$

Then the AOFM reads as

$$M_2^{(III)} = n^6 \left[p_{ef}^{(III)}\right]^2 \mathbf{r}^{-1}\left[v_{QT_1}(\Theta,\mathbf{j})\right]^{-3}. \quad (29)$$

As seen from Fig. 8, the EEC and the coefficient $M_2^{(III)}$ manifest petal-like dependences on the AW vector direction in the $XZ$ plane. Moreover, the EEC surface contains parts that differ by their signs. As with the AO interactions in amorphous media, the AO interaction of the type considered here for the cubic crystals manifests relatively small AOFM. It reaches its maximum $M_2^{(III)} = 0.63\times 10^{-15}\,\text{s}^3/\text{kg}$ for KBr crystals at $\mathbf{j} = 45\,\text{deg}$ and $\Theta = 65\,\text{deg}$ (see Fig. 9). The anisotropy of AOFM is mainly influenced by the anisotropy of EEC.

As mentioned above, the type (IV) of AO interaction, the interaction of the QT$_1$ wave with the optical wave polarized along the $Y$ axis, can be implemented in the cubic crystals. After rotation of the AW vector by the angle $\Theta$, the following strain tensor components appear:

$$\tilde{e}_1 = e_5\sin 2\Theta;\quad \tilde{e}_3 = -e_5\sin 2\Theta;\quad \tilde{e}_5 = e_5\cos 2\Theta. \quad (30)$$

The corresponding change in the refractive index is given by

$$\Delta n = n^3\left(p'_{21}\tilde{e}_1 + p'_{23}\tilde{e}_3 + p'_{25}\tilde{e}_5\right)/2. \quad (31)$$

The coefficient $p'_{25}$ is equal to zero for all symmetry groups of the cubic system, and the relationship $p'_{21} = p'_{23} = p'_{12}$ fulfils for the groups m3m, 432 and $\bar{4}$3m. This leads to zero refractive index change in Eq. (31). On the contrary, we have $p'_{21} \neq p'_{23}$ for the symmetry groups 23 and m3. Below we will consider this kind of diffraction separately.

Now let us analyze the AO interaction of optical waves with the AW QT$_2$ $v_{12} = v_{QT_2}$ propagating along the $X$ axis and polarized along the $Y$ axis, which produces the strain tensor component $e_6$. Depending on the orientation of electric induction of the incident light wave ($D_2$ or $D_3 = D\sin\mathbf{q}_B$ and $D_1 = D\cos\mathbf{q}_B$), the types (V) or (VI) of the AO interaction become actual. In the case (V), the

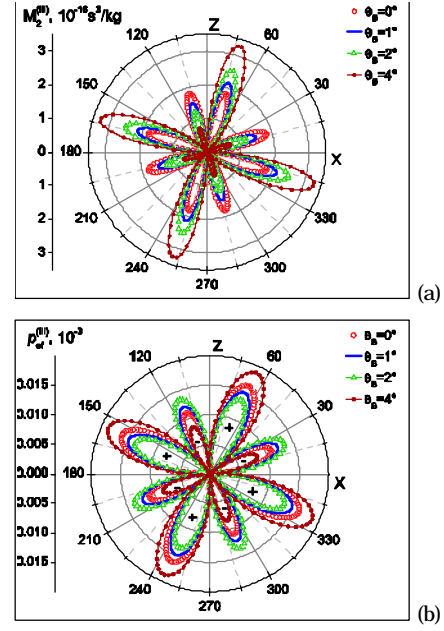

Fig. 8. Dependences of $M_2^{(III)}$ coefficient (a) and EEC (b) on the direction of AW vector (angle $\Theta$) at different values of $\mathbf{q}_B$ and at angle $\mathbf{j} = 0$ deg.

strain tensor includes the two components dependent upon the AW vector orientation:

$$\tilde{e}_6 = e_6\cos\Theta,\quad \tilde{e}_4 = -e_6\sin\Theta. \quad (32)$$

Then the increment of the optical frequency impermeability tensor and the corresponding increment of the refractive index can be written as

$$\Delta B_2 = -p_{16}e_6\cos\Theta, \quad (33)$$
$$\Delta n = -n^3 p_{16}e_6\cos\Theta/2, \quad (34)$$

with the EEC being equal to

$$p_{ef}^{(V)} = p_{16}\cos\Theta = -\sin 4\mathbf{j}\cos\Theta(p_{11}-p_{12}-2p_{44})/4. \quad (35)$$

As a result, the AOFM becomes as follows:

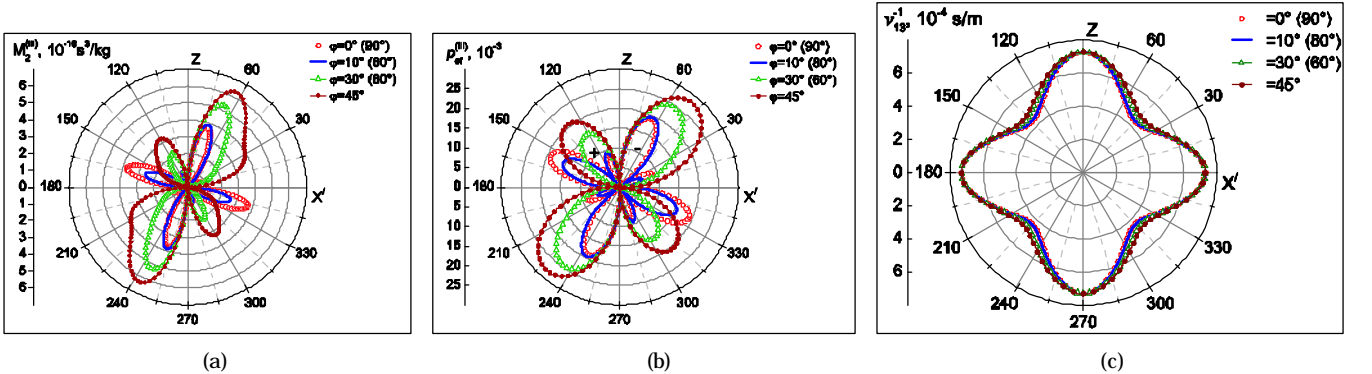

Fig. 9. Dependences of $M_2^{(III)}$ coefficient (a), EEC (b) and cross section of the acoustic slowness surface on the angle $\Theta$ for different orientations of AO

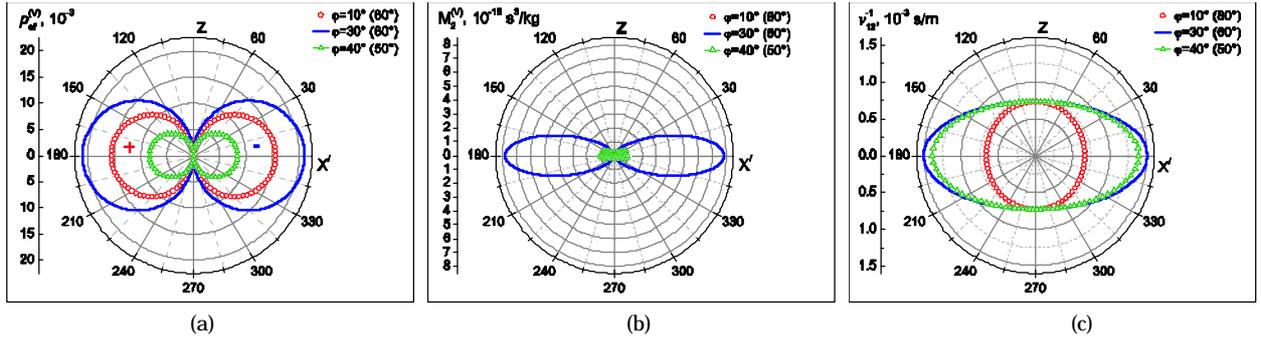

Fig. 10. Dependences of EEC (a), AOFM $M_2^{(V)}$ (b) and acoustic slowness $v_{12}^{-1}$ (c) on the angle $\Theta$ at different angles $\varphi$.

$$M_2^{(V)} = n^6 \left[ p_{ef}^{(V)} \right]^2 \rho^{-1} \left[ v_{QT_2}(\Theta,\varphi) \right]^{-3}. \qquad (36)$$

The dependences of the $M_2^{(V)}$ coefficient on the angles $\varphi$ and $\Theta$ are presented in Fig. 10. As seen from Fig. 10, the spatial dependence of AOFM in the $X'Z$ plane is similar to that typical for $p_{ef}$. Notice that at $\varphi = 0$, 90 and 45 deg the relation $p_{ef} = 0$ holds true. Then the AO diffraction does not occur. It is also interesting that the EEC surface consists of parts with the opposite signs. The AOFM becomes the highest for the direction where the AW is the slowest ($M_2^{(V)} = 7.82 \times 10^{-15}$ s³/kg at $\varphi = 30$ or 60 deg and $\Theta = 0$ deg).

Now consider the type (VI) of AO interaction when the polarization of incident and diffracted optical waves belongs to the $X'Z$ plane. The increment of the refractive index induced by the strains (34) due to the AW is given by

$$\Delta n = -n^3 p_{16} e_6 \cos\Theta \cos^2 \theta_B / 2. \qquad (37)$$

It is readily seen that Eq. (37) is reduced to Eq. (35) at small $\theta_B$ values. The same is true for the EEC (see Fig. 11,a) and the AOFM:

$$p_{ef}^{(VI)} = \left[ \sin 4\varphi \, (p_{11} - p_{12} - 2p_{44})/4 \right] \cos\Theta \cos^2 \theta_B, \qquad (38)$$

$$M_2^{(VI)} = n^6 \left[ p_{ef}^{(VI)} \right]^2 \rho^{-1} \left[ v_{QT_2}(\Theta,\varphi) \right]^{-3}. \qquad (39)$$

Hence, the AO interactions of the types (VI) and (V) are the same at small enough Bragg angles. The EEC and the AOFM (see Fig. 11) are also the same.

Following from our analysis of anisotropy of the AOFM for KBr crystals, we conclude that the highest $M_2$ value ($12.8 \times 10^{-15}$ s³/kg) is reached for the type (II) of AO interaction, the interaction of the incident optical wave polarized in the $XZ$ plane (i.e., in the crystallographic plane $ac$) with the longitudinal AW propagating along the bisector of the $X$ and $Z$ axes (or the crystallographic axes $a$ and $c$). This $M_2$ value is due to a slowness of the corresponding AW rather than due to elastooptic anisotropy. Notice that, owing to a high symmetry of the optical material, the same conclusion remains true for the other crystallographic planes.

Now let us consider the AO interaction in the crystals that belong to the point symmetry groups 23 and m3. As already mentioned, the elastooptic tensor for these groups consists of the four independent coefficients ($p_{11} = p_{22} = p_{33}$, $p_{44} = p_{55} = p_{66}$, $p_{12} = p_{23} = p_{31}$, and $p_{21} = p_{13} = p_{32}$). We will perform our numeric analysis for KAl(SO$_4$)$_2 \times 12$H$_2$O crystals belonging to the group m3 [22]. The refractive index, the elastooptic coefficients, the density and the elastic stiffness coefficients for these crystals are as follows: $n = 1.4564$, $p_{11} = 0.199$, $p_{12} = 0.260$, $p_{21} = 0.265$, $p_{44} = -0.005$ (at $\lambda = 589$ nm), $\rho = 1753$ kg/m³, $C_{11} = 24.65$ GPa, $C_{12} = 10.21$ GPa, and $C_{44} = 8.67$ GPa [22]. One can assume that the main peculiarities of AO interactions in the crystals with the symmetries 23 and m3 should be the same as those for the groups m3m, 432 and $\bar{4}$3m. Therefore we will present only relations for the EECs (respective AOFM are determined by the relations 24, 26, 29), which are more complicated then those analyzed before:

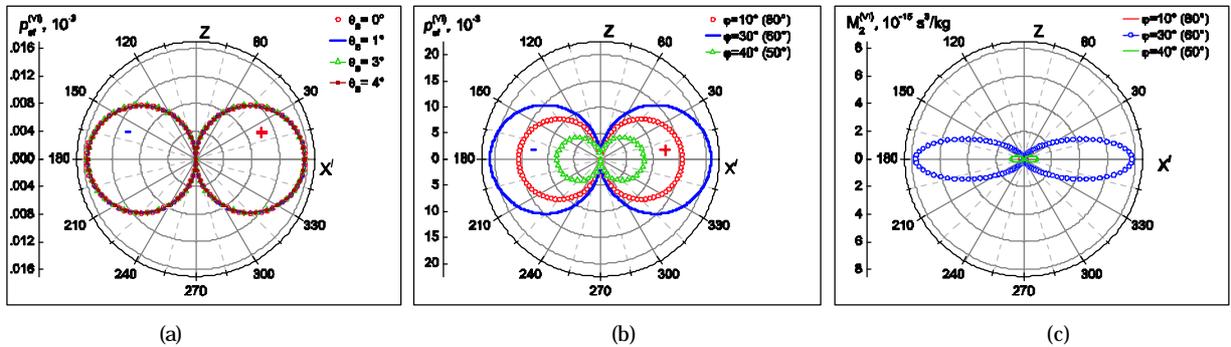

Fig. 11. Dependence of EEC on the angle $\Theta$ for different Bragg angles (a) and for different $\varphi$ angles at $\theta_B = 4$ deg (b); dependence of $M_2^{(VI)}$ coefficient on the angle $\Theta$ for different $\varphi$ angles at $\theta_B = 4$ deg (c).

$$p_{ef}^{(I)} = \left[\begin{array}{l} p_{12}\sin^4 j + p_{21}\cos^4 j + \\ +(p_{11}-2p_{44})\sin^2 2j /2 \end{array}\right]\cos^2\Theta + p_{12}\sin^2\Theta, \quad (40)$$

$$p_{ef}^{(II)} = \cos^2(q_B+\Theta)\times$$
$$\times\left\{\left[\begin{array}{l}p_{11}(\cos^4 j +\sin^4 j)+\\ \sin^2 2j\,(p_{12}+p_{21}+4p_{44})/4\end{array}\right]\cos^2\Theta+\right.$$
$$\left.+p_{21}\sin^2\Theta\right\}, \quad (41)$$
$$+\sin^2(q_B+\Theta)(p_{12}\cos^2\Theta+p_{11}\sin^2\Theta)-$$
$$-p_{44}\sin 2(q_B+\Theta)\sin 2\Theta/2$$

$$p_{ef}^{(III)} = \cos^2(q_B+\Theta)\sin 2\Theta\times$$
$$\times\left[\begin{array}{l}p_{11}(\cos^4 j +\sin^4 j)-p_{21}+\\ +\sin^2 2j\,(p_{12}+p_{21}+4p_{44})/4)\end{array}\right]+ \quad (42)$$
$$+\sin^2(q_B+\Theta)\sin 2\Theta(p_{12}-p_{11})+$$
$$+p_{44}\sin 2(q_B+\Theta)\cos 2\Theta$$

The type (IV) of interaction will be analyzed further on. Finally, for the types (V) and (VI) of AO interaction we have respectively

$$p_{ef}^{(V)} = -\frac{1}{2}\sin 2j\,\cos\Theta\left[\begin{array}{l}(p_{11}-2p_{44})\cos 2j +\\ +p_{12}\sin^2 j - p_{21}\cos^2 j\end{array}\right], \quad (43)$$

$$p_{ef}^{(VI)} = \frac{1}{2}\sin 2j\,\cos\Theta\cos^2 q_B\left[\begin{array}{l}(p_{11}-2p_{44})\cos 2j +\\ +p_{21}\sin^2 j - p_{12}\cos^2 j\end{array}\right], \quad (44)$$

Dependences of the AOFM on the AW vector direction calculated for the $KAl(SO_4)_2\times 12H_2O$ crystals with using of Eqs.(24, 26, 29, 36, 39) for the types of AO interaction mentioned above are presented in Fig. 12.

The cross sections of the surfaces corresponding to anisotropy of the AOFM reveal a two-fold symmetry instead of four-fold one peculiar for the crystals of the groups m3m, 432 and $\bar{4}3m$. Moreover, these cross sections are slightly rotated around the $Y$ axis. The acoustic slowness $v_{11}^{-1}$ for $KAl(SO_4)_2\times 12H_2O$ crystals is almost independent on the angle $\Theta$ (see Fig. 13,a) and, as a result, the anisotropies of $M_2^{(I)}$ and $M_2^{(II)}$ are determined by the EEC anisotropy (cf. with Fig. 14,a and Fig. 14,b). The maximum AOFM for the type (I) of interaction (see Fig. 12,a) is equal to $7.3\times 10^{-15}$ s$^3$/kg . It is reached at $j$ and $\Theta$ equal to 0 or 180 deg. The maximum figure of merit for the type (II) of interaction ($5.4\times 10^{-15}$ s$^3$/kg) is observed at $j=45$ and 135 deg, and $\Theta=137$ and 317 deg. The AOFM for the type (III) of interaction is small, being equal only to $0.16\times 10^{-15}$ s$^3$/kg (at $\Theta=62$ and 242 deg , and $j=45$ or 135 deg ). The anisotropy of $M_2^{(III)}$ is determined only by anisotropy of the EEC (cf. Fig. 12,c, Fig. 13,b and Fig. 14,c), since the parameter $v_{13}^{-1}$ is almost independent on the angle $\Theta$. For the types (V) and (VI) of interaction, the AOFM is also small ($M_2^{(V)}=0.11\times 10^{-15}$ s$^3$/kg at $\Theta=0$ deg and $j=20$ deg, and $\Theta=177$ or 357 deg and $j=110$ deg, respectively). The anisotropy of AOFMs in the two latter cases is determined only by anisotropy of the elastooptic coefficients (see Fig. 12,d, e, Fig. 13,c and Fig. 14,d, e). Again, $v_{12}^{-1}$ does not depend on the angle $\Theta$ for $KAl(SO_4)_2\times 12H_2O$ crystals. The EEC for the types (V) and (VI) of AO interaction is positive at $-45\deg<j<45\deg$, $135\deg<j<225\deg$ and $270\deg<\Theta<90\deg$, and negative at $90\deg<\Theta<270\deg$. The opposite situation occurs at $45\deg<j<135\deg$ and $225\deg<j<315\deg$ (see Fig. 14,d, e).

Let us finally consider the type (IV) of AO interaction, i.e. the interaction of the AW $QT_1$ with the optical wave polarized parallel to the $Y$ axis. As shown above, the strain tensor components are determined by Eq. (30) and the changes of the refractive index by Eq. (31). Then we obtain the EEC and the AOFM:

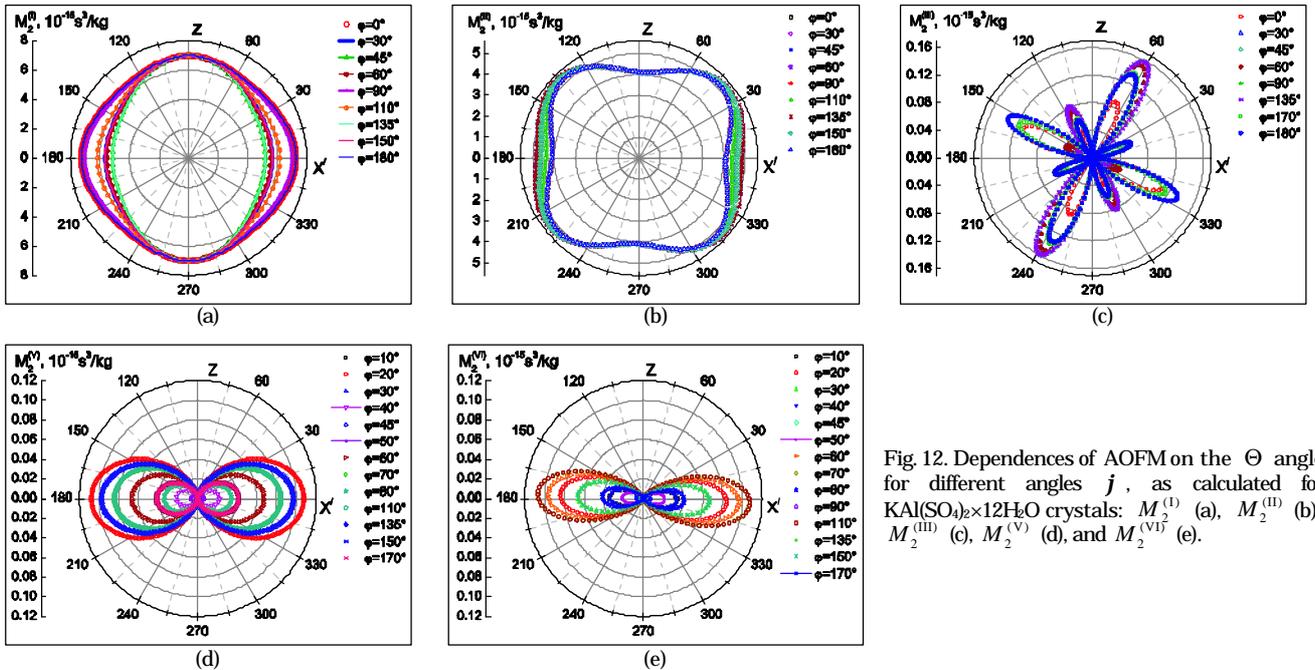

Fig. 12. Dependences of AOFM on the $\Theta$ angle for different angles $j$, as calculated for $KAl(SO_4)_2\times 12H_2O$ crystals: $M_2^{(I)}$ (a), $M_2^{(II)}$ (b), $M_2^{(III)}$ (c), $M_2^{(V)}$ (d), and $M_2^{(VI)}$ (e).

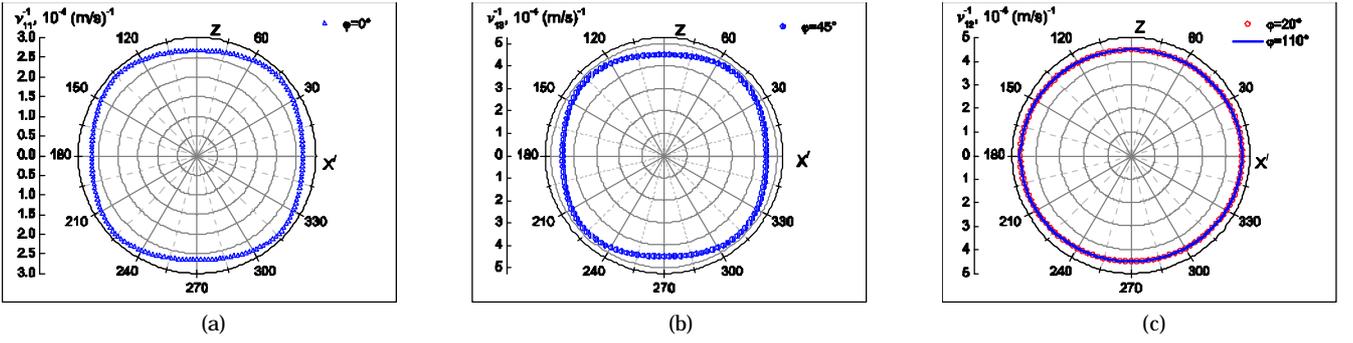

Fig. 13. Dependences of acoustic slownesses $v_{11}^{-1}$ (a), $v_{13}^{-1}$ (b) and $v_{12}^{-1}$ (c) on the angle $\Theta$ at the angles $\varphi$ corresponding to the maximal AOFMs.

$$p_{ef} = \begin{pmatrix} p_{21}\cos^4\varphi + p_{12}\sin^4\varphi - p_{12} + \\ + (p_{11} - p_{44})\sin^2 2\varphi/2 \end{pmatrix} \sin 2\Theta, \quad (45)$$

As seen from Fig. 15,a, the anisotropy of the AOFM is determined by anisotropy of the EEC (Fig. 15,b) rather than the acoustic slowness (cf. with Fig. 13,b). The maximum value $M_2^{(IV)} = 0.43 \times 10^{-15}$ s$^3$/kg occurs at $\Theta = 45$, 135, 225 and 315 deg, and $\varphi = 45$ and 135 deg. Notice that $M_2^{(IV)}$ at $\varphi = 0$ and 180 deg is equal to zero. Hence, the anisotropy of the AOFM for the KAl(SO$_4$)$_2 \times$12H$_2$O crystals, which manifest insignificant acoustic anisotropy, is determined by the anisotropy of EEC.

### 3. Concluding Remarks

In the present work we have suggested the analytical approach for analyzing the anisotropy of AOFM, which is based upon symmetry conditions for the acoustic and optical properties and their tensorial description. The relations for the EECs and the AOFMs are derived for all possible types of AO interactions. As a result of our analysis concerned with the anisotropy of AO interactions in optically isotropic media, one can make the following conclusions. Only one type of AO interaction can exist in the isotropic liquid and gaseous materials. These materials are therefore characterized by a single value of AOFM, which does not depend on the interaction geometry. The isotropic amorphous solid-state materials are to be characterized by three different types of AO interactions and so by three AOFMs. The first two types and the type (III) of interaction correspond to AO diffraction respectively on the longitudinal and transverse waves. The peculiarity of the type (III) of interaction is a relatively small AOFM. The figures of merit corresponding to all of these three types of AO interaction do not reveal anisotropy. The crystals belonging to the cubic symmetry system are characterized by six different types of AO interactions and therefore by six different AOFMs. These types of interactions correspond to isotropic diffraction of orthogonally polarized incident optical waves on the QL, QT$_1$ and QT$_2$ AWs. It is shown that, depending on the AW vector orientation, all of the six AOFMs reveal a notable anisotropy. It is caused mainly by the anisotropy of AW velocities and, in the case of isotropic AW velocities, by the anisotropy of EECs.

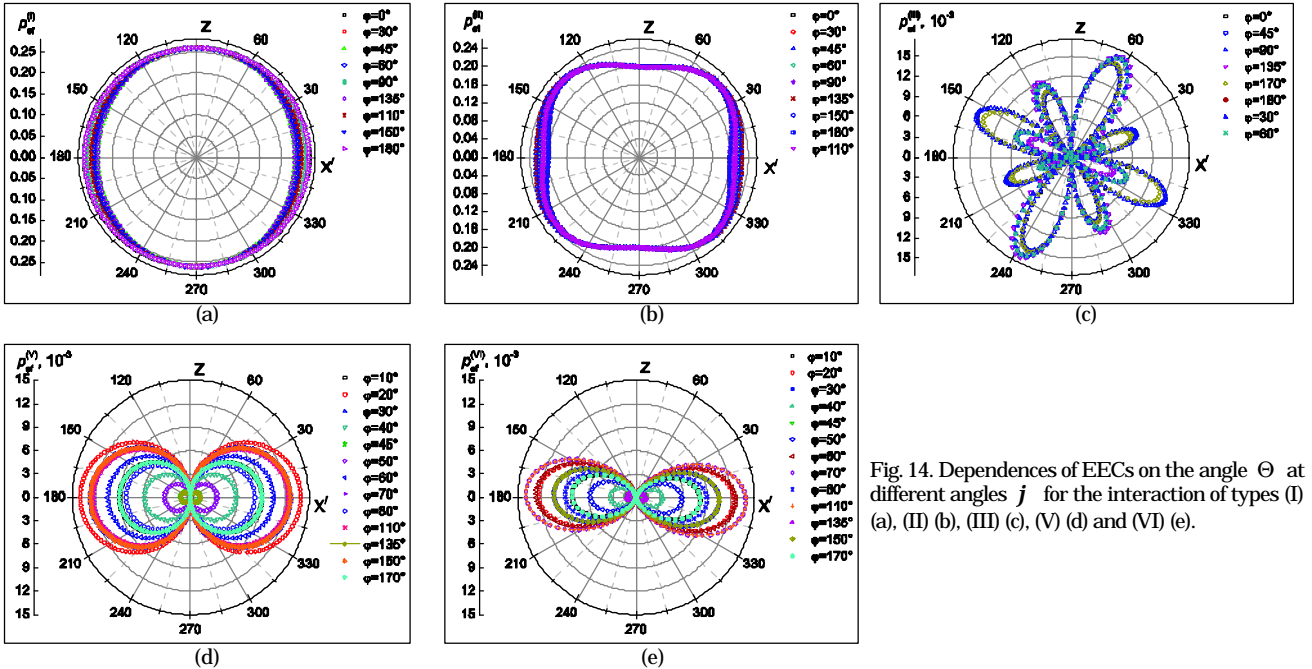

Fig. 14. Dependences of EECs on the angle $\Theta$ at different angles $\varphi$ for the interaction of types (I) (a), (II) (b), (III) (c), (V) (d) and (VI) (e).

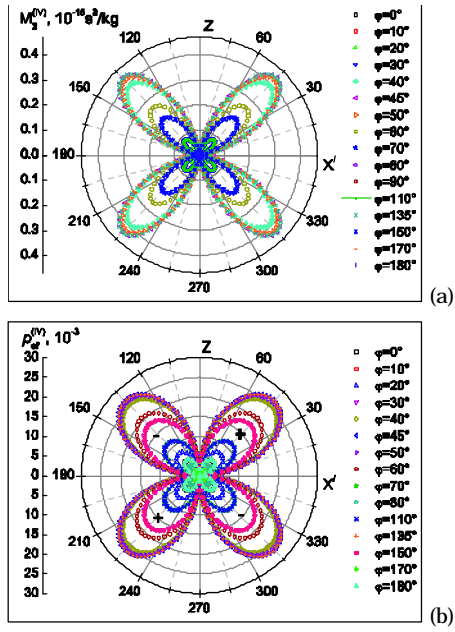

Fig. 15. Dependences of $M_2^{(IV)}$ (a) and EEC (b) on the angle $\Theta$ at different angles $\varphi$.

Our approach allows for finding the optimal geometry of AO interactions, which is characterized by the greatest AOFM for a certain material. The numeric analysis has been carried out for the cubic crystals KBr and KAl(SO$_4$)$_2$×12H$_2$O representing different subgroups of the cubic system. These crystals also differ by the anisotropy of AW velocities. Namely, KBr reveals a notable anisotropy of those velocities, whereas KAl(SO$_4$)$_2$×12H$_2$O is almost isotropic from this viewpoint. We have found that the highest $M_2$ for the KBr crystals ($12.8\times10^{-15}$ s$^3$/kg) is reached when the incident optical wave polarized in the crystallographic planes *ac*, *ab* or *bc* interacts with the longitudinal AW propagating along the *ac*, *ab*, or *bc* bisectors. The maximal $M_2$ value occurs due to significant slowness of the AWs participating in the AO interaction rather than because of the elastooptic anisotropy. In contrast, anisotropy of the AOFMs for the KAl(SO$_4$)$_2$×12H$_2$O crystals exists only due to anisotropy of the EECs, since the acoustic properties of these crystals are almost isotropic. Here the maximal AOFM ($7.3\times10^{-15}$ s$^3$/kg) is reached at $\varphi$ and $\Theta$ angles equal to 0 or 180 deg, respectively.

In the present work we have not taken into account the deviations from orthogonality (or longitudity) for the acoustic displacement vector with respect to the wave vector. This effect can induce additional components of the strain tensor and so affect the EEC and the AOFM. To our opinion, it would be reasonable to account for such small corrections when considering some specific practical cases.